\documentclass[preprint]{elsarticle}
\usepackage{graphicx,url,hyperref,microtype,subcaption,graphbox,amsmath}
\usepackage{latexsym,courier,upgreek,xcolor,rotating,ulem}
\usepackage{amssymb}
\usepackage{array}
\usepackage{paralist}
\usepackage[onehalfspacing]{setspace}
\hypersetup{colorlinks,citecolor=blue,linkcolor=red,urlcolor=green}
\usepackage[displaymath,mathlines]{lineno}
\usepackage[margin=2cm]{geometry}
\modulolinenumbers[5]
\biboptions{numbers,sort&compress}

\begin{document}

\begin{frontmatter}
  
  \title{Numerical Modeling of Solvent Diffusion through the Transition Metal Dichalcogenides based Nanomaterials}
  
  \author{Geetika Sahu}
  \ead{p20210061@hyderabad.bits-pilani.ac.in}
  \address{Dept. of Physics, Birla Institute of Technology \& Science Pilani, Hyderabad Campus, Secunderabad, Telangana 500078.}

\begin{abstract}

This article presents a numerical simulation of solvent diffusion in transition-metal dichalcogenide-based nanomaterials during solvothermal reaction, leading to layer exfoliation and, consequently, a reduction in the average nanoparticle size. By solving modified Fick’s law of diffusion and utilizing the dynamic bond percolation model, this study examines the evolution of a system of nanoparticles. During the simulation, the effects of key parameters, such as the diffusivity variable that determines the diffusion rate, and the number of iterations needed to achieve enhanced nanoparticle size uniformity, have been analyzed. To gain more insight into the size evolution of the nanoparticles, avalanche statistics, and fluctuations in the average nanoparticle size by Shannon entropy has been utilized. The size distribution observed for different diffusivity variables and iterations has also been studied, which predicts the probability of finding the nanoparticles of specific sizes within the system. A correlation between the iteration for maximum entropy and the minimum of relative change in particle size with iteration has been established, indicating that an iteration exists that takes the system towards saturation in terms of the average size of the nanoparticles. The numerical findings indicate that the experimental parameters, such as solvent selectivity and diffusivity, as well as reaction time, play significant roles in determining nanoparticle size and uniformity, thereby enhancing potential material applications.

\end{abstract}

\end{frontmatter}


\section{Introduction}

Transition metal dichalcogenides (TMDCs), a family of two-dimensional materials \cite{manzeli17} are versatile in terms of their electronic, optical, and catalytic properties \cite{Cao18,kumar25}. TMDCs are represented as $\mathrm{MX_2}$, where, M is the transition metal (such as molybdenum, tungsten) and X is the chalcogen (such as sulfur, selenium). It has emerged as a material inspired by the versatility of graphene and offers bandgap tunability based on particle size variation, achieved by scaling the material down to the zero-dimensional Quantum dots (QDs) \cite{uddin25}. Inter-layer van der Waals interactions and intra-layer covalent bonds characterize these QDs and exhibit fascinating optical properties such as size-dependent photoluminescence, tunable band gaps, and electronic properties as well as and enhanced surface activity due to the quantum confinement effect \cite{emil06, mansur10}. TMDC-based QDs offer promising applications such as in bio-imaging, optoelectronic, photo-catalysis, energy storage and nanocomposite based applications \cite{pratap25, haz22, sahu26}.

\begin{figure*} 
\centering
\includegraphics[width=0.75\textwidth]{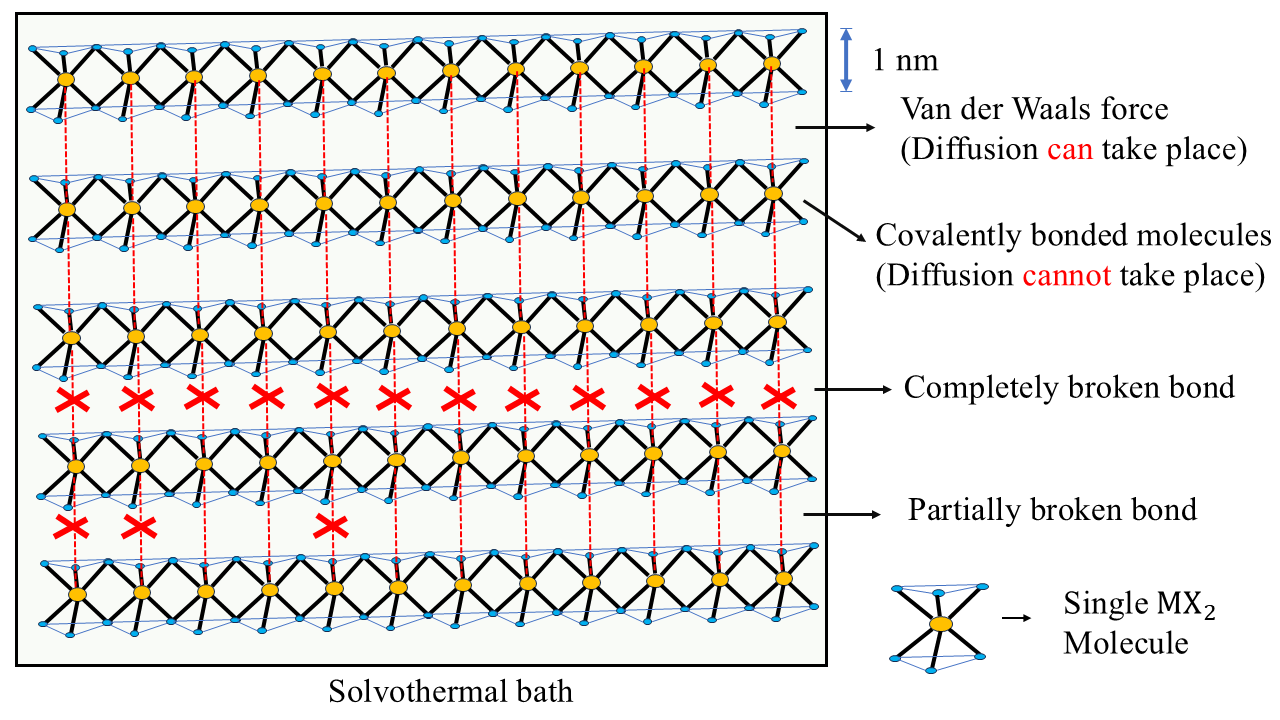}\ 
\caption{\footnotesize{Schematic diagram of a transition metal dichalcogenide-based nanoparticle in the solvothermal bath. The Figure represents van der Waals stacking of five layers of covalently bonded $\mathrm{MX_2}$ molecule. Due to the specific bonding interactions among molecules, solvent diffusion occurs only between layers not within the same layer. A red cross mark signifies a broken bond, which takes place when the solvent level exceeds the threshold of solvent level for that site. Therefore, a layer will undergo complete exfoliation once all the bonds within that layer are broken by solvent diffusion.}}
\label{fig_schematic}
\end{figure*}

One of the significant challenges in effectively utilizing the functionalities of QDs is achieving precise control over their size distribution during the synthesis process \cite{jana09, Zh15}. The size of the QDs is critical, as it directly influences their optical and electronic properties. Solvothermal methods and various solution-based exfoliation techniques are employed to effectively tune the size of QDs. These methods allow for careful tuning of the particle size, thereby enhancing the desired properties of the QDs. However, producing the QDs with a uniform size distribution using solvent assisted methods remains a complex task due to the presence of a number of experimental parameters involved in process, such as selection of solvent, concentration of the precursors, reaction time as well as temperature \cite{mehrer2007, sahu24,sahu25}. As per the experimental studies all these parameters independently play very important role in determining the evolution of nanoparticles \cite{becker11, rama18}. For example, the choice of solvent significantly affects the reaction rate. In addition to that, the factors such as solvent's pH and molecular size, which aid its intercalation into the host material are also important. Also, solvent selection should be done carefully so that the nanoparticle's internal structure is not damaged. Thus, all these parameters play a key role in determining the size evolution of nanoparticles. Hence, there is always a need of refinement and optimization of the reaction parameters in solvent assisted techniques \cite{kov21, dem08}.

During solvent assisted methods for the intercalation and exfoliation of layers in 2D TMDCs-based nanomaterials, the structure of the nanomaterials evolve due to its interaction with the solvent. Hence, for such system dynamic bond percolation model (DBPM) can be adapted \cite{harris86}. In principle, the diffusion of the solvent through the layered material changes the structure by deforming the stacking layers or inducing defects, hence, the system of particles is dynamic. This process leads to the exfoliation of layers and a reduction in the size of the nanoparticles \cite{ali22,wang22}.  

In this study, numerical simulation using Fick’s second law of diffusion has been employed to understand the evolution of a system of nanoparticles inside a solvothermal bath under constant temperature, pressure and concentration of the nanoparticles. The major focus is to tune the size of the nanoparticles and to achieve the uniformity in the size distribution of the QDs, hence optimizing the parameters involved in the diffusion of solvent through the material, such as the diffusivity of the solvent and the time taken (here taken in terms of iteration) to achieve the size uniformity in QDs. This diffusion driven framework gives a predictive pathway to understand how the average size of the nanoparticles changes by changing the solvent parameters, and hence monitoring the reaction using the rate of diffusion under solvent asssisted methods, particularly solvothermal synthesis route.
\begin{figure*} 
\centering
\includegraphics[width=0.5\textwidth]{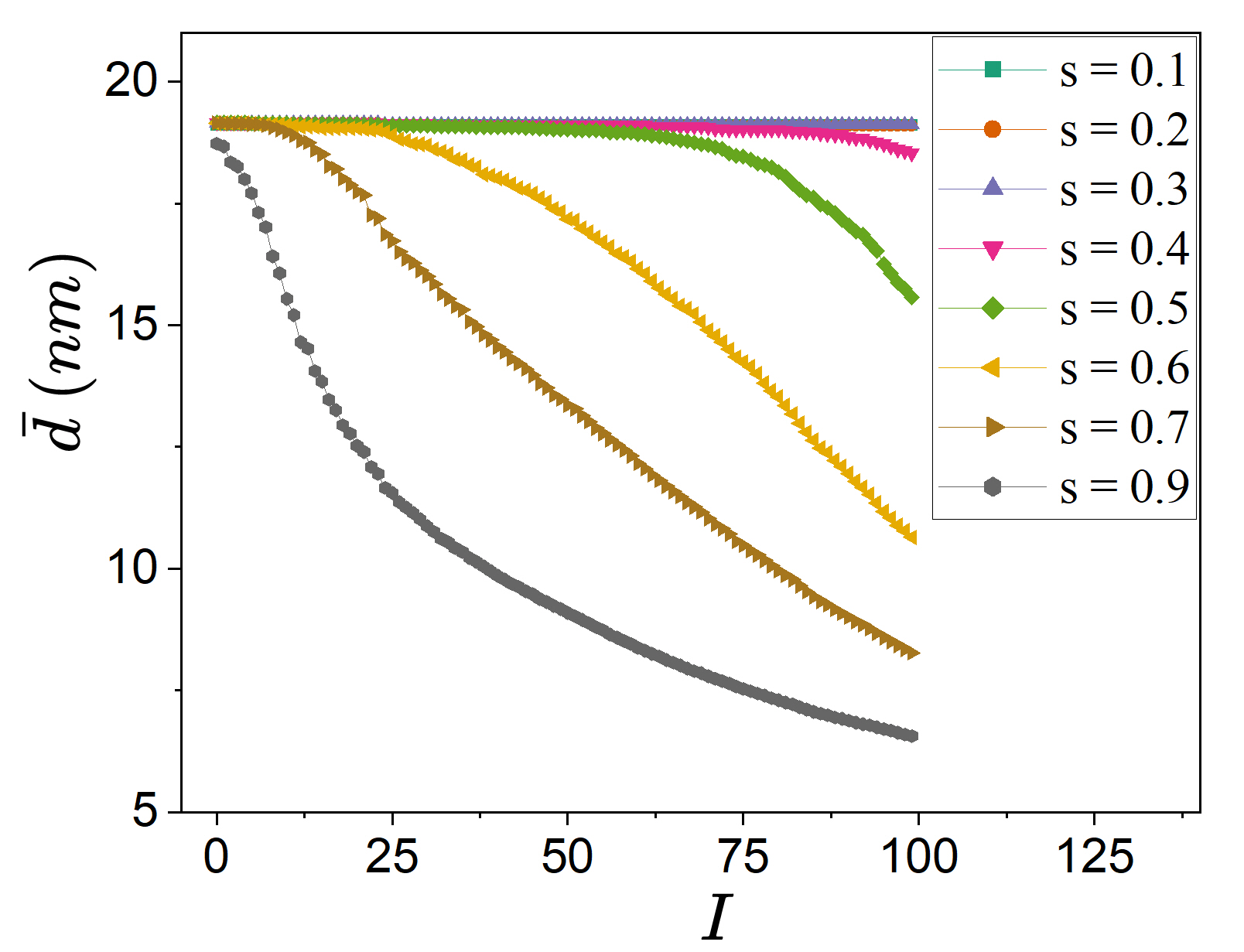} \ 
\caption{\footnotesize{ Variation in the average size ($\bar{d}$) of the nanoparticles with respect to the number of iterations $(I)$ for different diffusivity variables $(s)$ ranging from $s$ = 0.1 to 0.9. The results demonstrate that increasing the diffusivity variable enables solvent diffusion to overcome van der Waals forces between nanoparticle layers within a few iterations, thereby reducing $\bar{d}$. For smaller $s$ values, the average size remains constant, suggesting that the solvent's strength is insufficient to overcome the interlayer van der Waals forces.}}
\label{fig_1}
\end{figure*}

\section{Numerical system set-up and Flooding experiment}

To simulate and study the formation of smaller nanoparticles from larger ones through solvent intercalation between the layers of the nanoparticles via diffusion, the dynamic bond percolation model \cite{harris86, druger83} has been utilized. The dynamic bond percolation model can be utilized in this study due to the structural disordness in the transition-metal dichalcogenide (TMDC)-based nanoparticles. TMDC-based nanoparticles exhibit a sandwich structure, in which weak van der Waals forces hold together the layers and strong covalent bonds link the molecules within each layer \cite{samy21, guo20}. Due to this reason, the solvent diffusion can only take place in between the layers but not within a particular layer. A schematic representing a particle is shown in the figure \ref{fig_schematic}. The figure represents five stacked layers of $\mathrm{MX_2}$ molecule, having the uniformity in the width of each layer. The thickness of each layer is taken as $1\ nm$. The size of each nanoparticle can be calculated as:

\begin{align}
    d = \sqrt{\text{$t$} \times \text{$w$}}
\end{align}

where, $d$ is the size of the particle in $nm$, $t$ represents the total thickness of the particle, which is the sum of the thicknesses of each layer, and $w$ denotes the total width constituted by all the intact layers.  At the start of the simulation, it has been assumed that the nanoparticles have the same number of layers, with variations in their widths. The variation in width of the layers spans from $1$ to $15\ nm$ in this study.

The diffusion process has been realized using modified Fick’s second law of diffusion, given by \cite{wang22, xu92} : 

\begin{equation} \label{eq1}
\begin{split}
    P_{i,j}^{n+1} = P_{i,j}^n + s\ (W_{i+1,j}P_{i+1,j}^n+ W_{i-1,j}P_{i-1,j}^n\\ +W_{i,j+1}P_{i,j+1}^n + W_{i,j-1}P_{i,j-1}^n  - 4W_{i,j}P_{i,j}^n )
\end{split}
\end{equation}

In the equation above, $P_{i,j}^{n+1}$ denotes the updated solvent level based on the previous solvent level $P_{i,j}^{n}$ for a site $(i,j)$ during the diffusion process. $s$ represents the diffusivity variable, which influences the rate of diffusion. $s$ will vary depending upon the selection of solvents or by varying the reaction parameters such as temperature, concentration etc., for a particular solvent \cite{Lee05,chan14}. To introduce heterogeneity or defects in terms of material properties, a parameter $W$ \cite{wang22, xu92} has been introduced, which refers to the diffusion path or hopping rate associated with all sites. $W$ indicates the probability of diffusion between adjacent sites, generated randomly from uniform distribution within the range [0.1, 0.9], before the simulation begins. When $W$ is zero, the path between the adjacent sites is blocked, implying that even if there is a significant difference in the solvent level between them, diffusion will not occur.

Equation \ref{eq1} is derived using the finite difference method to solve modified Fick’s law of diffusion \cite{rahaman15}:
\begin{equation} \label{eq2}
   \frac{\partial P}{\partial t}  = D \left( \frac{\partial ^2P'}{\partial x^2} + \frac{\partial ^2P'}{\partial y^2} \right)
\end{equation}

where, D is diffusion coefficient having the unit ${m^2}/s$, and $P' = PW$, since $W$ has a random map in space that does not change with time.

Using space discretizations, $\frac{\partial ^2P'}{\partial x^2}$ and $\frac{\partial ^2P'}{\partial y^2}$ can be approximated for a site (i,j) as,

\begin{align}\label{eq3}
    \frac{\partial ^2P'_{i,j}}{\partial x^2} &\approx \frac{P_{i+1,j}W_{i+1,j} - 2P_{i,j}W_{i,j} + P_{i-1,j}W_{i-1,j}}{(\Delta x)^2}\\
    \frac{\partial ^2P'_{i,j}}{\partial y^2} &\approx \frac{P_{i,j+1}W_{i,j+1} - 2P_{i,j}W_{i,j} + P_{i,j-1}W_{i,j-1}}{(\Delta y)^2}
\end{align}

and using time discretization, $ \frac{\partial P_{i,j}}{\partial t}$ is given by,

\begin{equation} \label{eq5}
   \frac{\partial P_{i,j}}{\partial t} \approx \frac{P_{i,j}^{n+1} - P_{i,j}^n}{\Delta t}
\end{equation}

Substituting in equation \ref{eq2}, we get

\begin{align}  \label{eq6}
    \frac{P_{i,j}^{n+1} 
    - P_{i,j}^n}{\Delta t} = & D\Bigg[ \frac{P_{i+1,j}W_{i+1,j} - 2P_{i,j}W_{i,j} + P_{i-1,j}W_{i-1,j}}{(\Delta x)^2} \nonumber \\  & + \frac{P_{i,j+1}W_{i,j+1} - 2P_{i,j}W_{i,j} + P_{i,j-1}W_{i,j-1}}{(\Delta y)^2} \Bigg]   
\end{align}

\begin{align}  \label{eq7}
    P_{i,j}^{n+1} = & P_{i,j}^n + s_x (P_{i+1,j}W_{i+1,j} - 2P_{i,j}W_{i,j} + P_{i-1,j}W_{i-1,j}) \nonumber  \\  & + s_y (P_{i,j+1}W_{i,j+1} - 2P_{i,j}W_{i,j} + P_{i,j-1}W_{i,j-1})
\end{align}

Here, s is the diffusivity variable, given by 

\begin{align} 
   s_x = \frac{D \Delta t}{(\Delta x)^2}, 
   s_y = \frac{D \Delta t}{(\Delta y)^2}
\end{align} \label{eq8}

Assuming discretizations in x and y direction are same. Hence, $\Delta x = \Delta y$, so, $s_x = s_y = s$. Hence, we get
\begin{equation} 
\begin{split}
     P_{i,j}^{n+1} = P_{i,j}^n + s (W_{i+1,j}P_{i+1,j}^n+ W_{i-1,j}P_{i-1,j}^n + \\W_{i,j+1}P_{i,j+1}^n + W_{i,j-1}P_{i,j-1}^n  - 4W_{i,j}P_{i,j}^n )
\end{split}
\end{equation}

This equation will be used to study the diffusion of solvent across the system of nanoparticles. Inside a solvothermal bath, there is a possibility of diffusion of solvents from all the sides in a nanoparticle. Hence, at the boundaries of the nanoparticles $P_{i,j}$ is set to be $1$. As diffusion progresses, a bond is considered broken once the threshold $P_{i,j} = 0.9$ is exceeded. When all the bonds within a particular layer are broken, this indicates that the single particle has split into two smaller particles.

For the numerical simulation, 50 configurations have been taken each consisting of 50 layers, representing total thickness of the particles as 50 nm before the simulation starts. The widths of these layers are not uniform and display randomness from $1$ to $15\ nm$ chosen from the uniform distribution, which helps to more accurately mimic real particles in the model. Different values of $s$ ranging from 0.1 to 0.9 are taken, and the change in system has been studied over 100 iterations (denoted as $I$). To gain insight into the overall behavior of the system with variation in the parameters $s$ and $I$, the average size $(\bar{d})$ of the nanoparticles have been calculated. This is accomplished by taking the mean of the $d$ across all the configurations. 
\begin{figure*}[ht]
\centering
\includegraphics[width=0.8\textwidth]{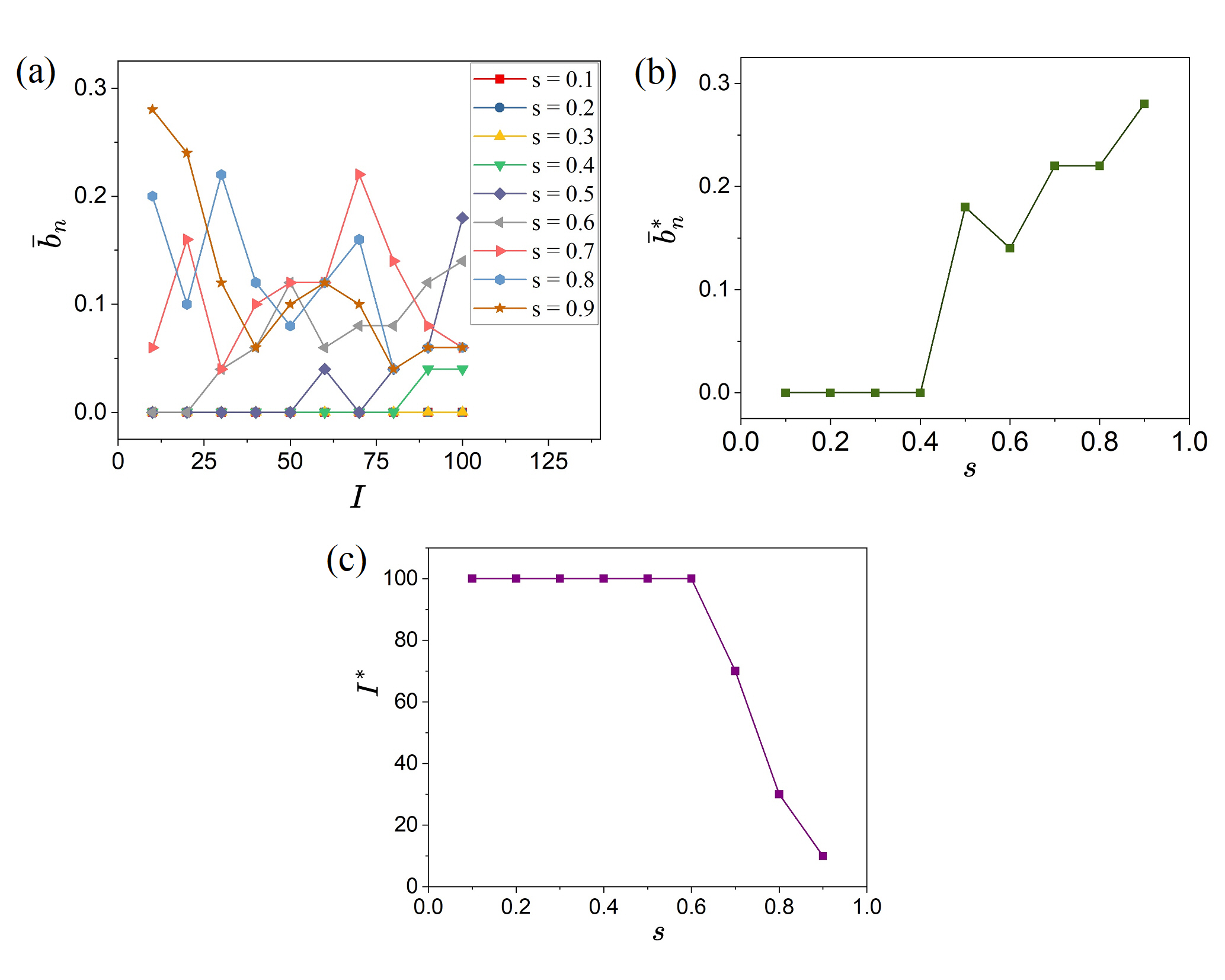} \ 
\caption{\footnotesize{(a) Variation in average number of new bonds broken ($\bar{b}_n$) with respect to $I$ for different values of $s$, indicating that for smaller $s$ values no new bonds are breaking even at 100th iteration. In contrast, intermediate and higher $s$ values lead to a noticeable increase in the number of new bonds broken within just a few iterations during the diffusion simulation. (b) The maximum number of new bonds broken ($\bar{b}_n^*$) for different diffusivity variables. (c) Iteration for the maximum number of new bonds broken represented as $I^*$ with respect to different diffusivity variables.}}
\label{fig_2}
\end{figure*}

\section{Results and Discussion}

\begin{figure*}
\centering
\includegraphics[width=1\textwidth]{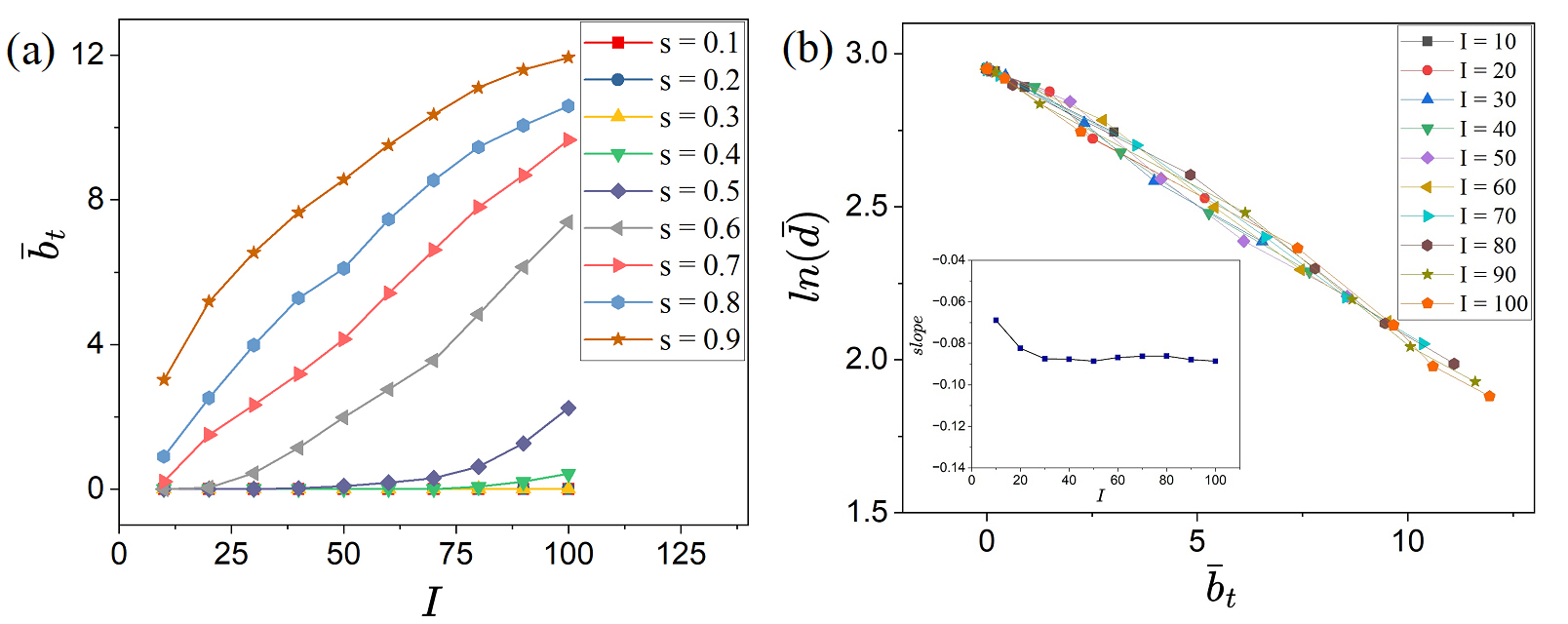} \ 
\caption{\footnotesize{(a) Variation in average number of total bonds broken ($\bar{b}_t$) with respect to $I$ for different diffusivity variables. For higher and intermediate $s$, $\bar{b}_t$ is an increasing function with $I$, whereas for small $s$, $\bar{b}_t$ does not show any increment till $I\ =\ 100$. (b) Variation in the $ln(\bar{d})$ with respect to $\bar{b}_t$ for different iterations across all $s$ ranging from 0.1 to 0.9. The plot indicates that the $\bar{d}$ of the nanoparticles reduces exponentially with increasing $\bar{b}_t$ and is lowest for the highest value of $s$ observed over all the iterations. Inset shows variation of slope obtained from $ln(\bar{d})$ vs $\bar{b}_t$, with respect to $I$ indicating a saturation in the slope values. The saturation shows that the reduction of $\bar{d}$ in relation to $\bar{b}_t$ with increasing $s$ values is consistent across all $I$ values.}}
\label{fig_3}
\end{figure*}

\subsection{Size evolution of the nanoparticles}

Figure \ref{fig_1} shows the variation in $\bar{d}$ of the nanoparticles with the number of iterations for different values of diffusivity variable. The plot shows that for smaller values of $s$, specifically $s$ = 0.1, 0.2 and 0.3, $\bar{d}$ remains constant at around $18\ nm$ throughout 100 iterations. In contrast, for intermediate values of $s$, $\bar{d}$  gradually decreases as the number of iterations increases. For instance, for $s$ = 0.5, $\bar{d}$  starts to decrease from $18\ nm$ around the intermediate values of $I$ and drops to $15\ nm$ by the 100th iteration.
For very high values of the diffusivity variable, specifically $s = 0.8$  and $s = 0.9$, it can be observed that the $\bar{d}$ significantly decreases even for a small number of iterations. This is particularly evident in the plot for $s = 0.9$, where within 25 iterations, $\bar{d}$ reduces to $11.6\ nm$ and continues to decrease to $6.5\ nm$ by $I=100$. This result indicates that for a higher rate of diffusion the possibility of intercalation of solvents in between different layers of nanoparticle increases, which leads to reduction in their $\bar{d}$. This reduction in $\bar{d}$ significantly depends on number of iterations, which can be linked to the experimental reaction time in the hydrothermal reaction \cite{john18,sahu24}. 

It can be observed that for higher $s$ and corresponding higher $I$, the rate at which $\bar{d}$ is reducing with respect to $I$ is also reducing, slowly taking the system towards a saturation. The reduction in rate of change of $\bar{d}$ with respect to $I$ indicates that more uniformity in the particle is being achieved with increase in $I$ values for higher $s$ values. This phenomenon occurs because, after a significant number of larger particles have been broken down into smaller ones, there are fewer bonds available to break. Beyond a certain iteration value, only the smallest size achievable $\bar{d}$ for the specific setup will remain in the system. This represents the saturation of the average particle size in that system. This has been discussed in details in the further section.

\subsection{Avalanche Statistics}
\begin{figure*} [ht]
\centering
\includegraphics[width=0.9\textwidth]{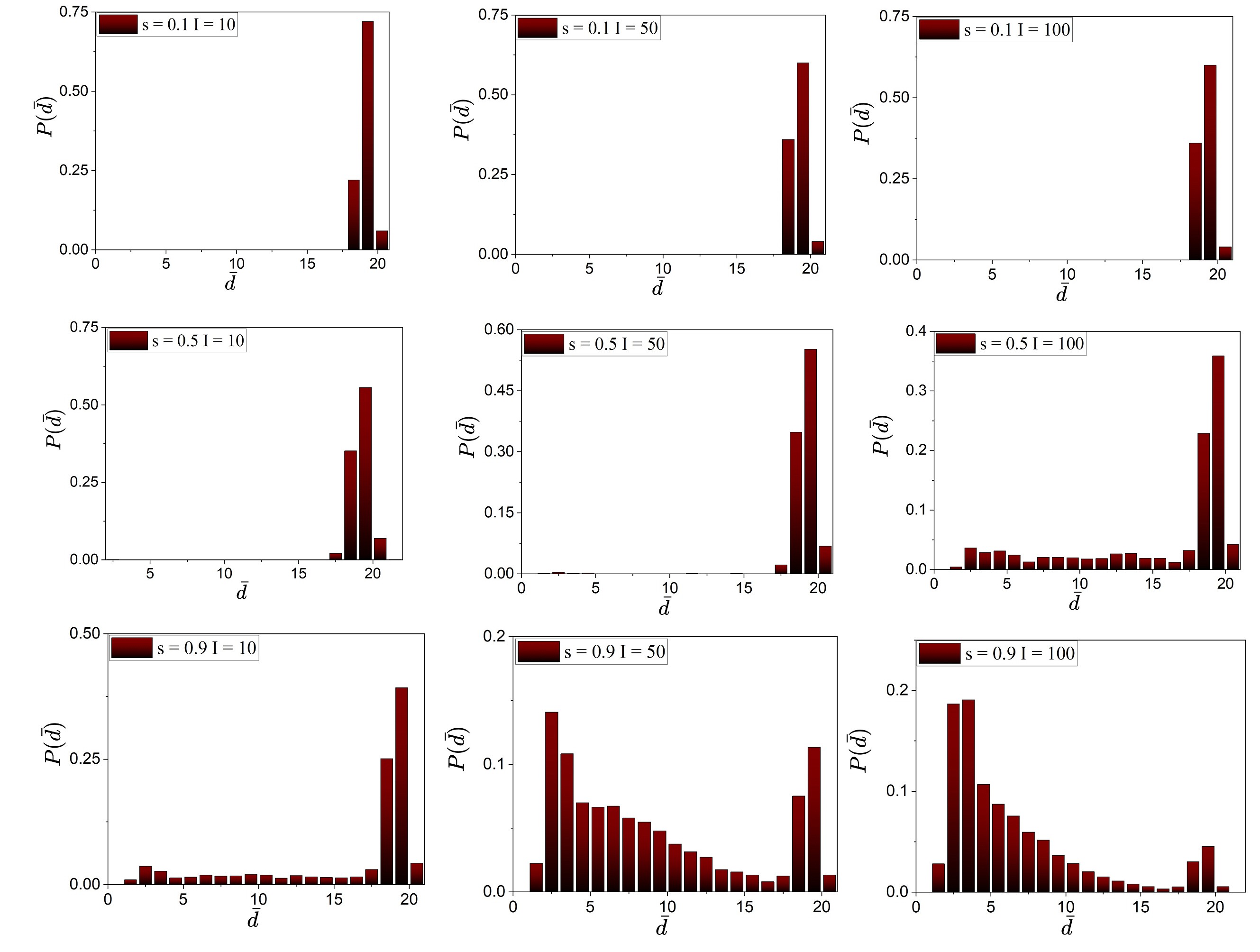} \\
\caption{\footnotesize{Size distribution of particles at varying diffusivity variable for three different iterations. The distributions are presented for s = 0.1, 0.5 and 0.9, at I = 10, 50 and 100. For $s\ =\ 0.1$, the first three plots indicate that the average size $\bar{d}$ of nanoparticles remains constant around $18\ nm$ across all iterations. The subsequent plots for $s\ =\ 0.5$ reveal an increasing probability of smaller particles emerging in the system at $I\ =\ 50$, which is further increasing by $I\ =\ 100$. The last row indicates the size distribution plots for $s\ =\ 0.9$, where a significant non-uniformity in the $\bar{d}$ of the nanoparticles can be observed within $I\ =\ 10$, which is increasing with an increase in number of iterations.}}
\label{fig_4}
\end{figure*}

To understand the particle size evolution, a systematic study of the average number of new bonds broken ($\bar{b}_n$) at each iteration of the diffusion process, as well as the average cumulative number of broken bonds ($\bar{b}_t$) until all iterations have been performed. From figure \ref{fig_2}(a), it is evident that $\bar{b}_n$ remains constant at zero throughout all $I$ for $s= 0.1,\ 0.2$ and $0.3$, indicating no new bonds have been broken throughout the diffusion process at any value of $I$. In the case of intermediate values of $s$, $\bar{b}_n$ begins to exhibit an increase around the intermediate values of $I$. Conversely, for higher values of $s=0.6, 0.7, 0.8$ and $0.9$, $\bar{b}_n$ starts to increment right from the beginning of the diffusion process. This indicates that for higher values of $s$ there is faster diffusion of solvents. The diffusion process combined with the material property given by the diffusion path $(W)$ leads to the delamination of layers resulting in the reduction of $\bar{d}$. There is another interesting point to be noted here in the trend observed in figure \ref{fig_2}(a). In case of $s=0.9$, it is observed that the $\bar{b}_n$ initially exhibits a very high value of $0.28$ indicating, 28\% of the total bonds have already been broken by $I\ =\ 10$. This value subsequently decreases with an increasing number of iterations. This decreasing trend can be attributed to the diffusion of the solvent and the availability of bonds to be broken; as the diffusion of solvent proceeds towards the higher values of $I$, the availability of bigger particles to undergo delamination reduces, resulting in lower $\bar{b}_n$ values in the later iterations. It should also be noted here that, as the size of the nanoparticles is reduced to a great extent, the surface area to volume ratio significantly increases, and hence higher surface energy \cite{wu25, coleman13}. For the diffusion process to take place further for the delamination, it has to overcome the surface energy contributed by the unsatisfied surface bonds, also known as dangling bonds in TMDCs-based QDs \cite{loh15}. In the synthesis process, surfactants play a crucial role by balancing the dangling bonds, which helps to lower the surface energy of the nanoparticles \cite{loh15, abr22}. As a result, when the reaction rate is increased by increasing the temperature or by using a particular solvent, the particles can be further fragmented into smaller sizes. In the simulation, since there are smaller particles in the system after $I\ =\ 10$, they will have an increased surface area to volume ratio. The competition between the surface energy and the diffusion process for delamination of layers, which further reduces particle size, will also increase \cite{wu25, coleman13}. However, as indicated by figure \ref{fig_1} at higher $I$, the $\bar{d}$ further decreases, indicating that $s$ is sufficiently high to delaminate and break the particles into smaller sizes.  
\begin{figure*}[ht]
\centering
\includegraphics[width=0.85\textwidth]{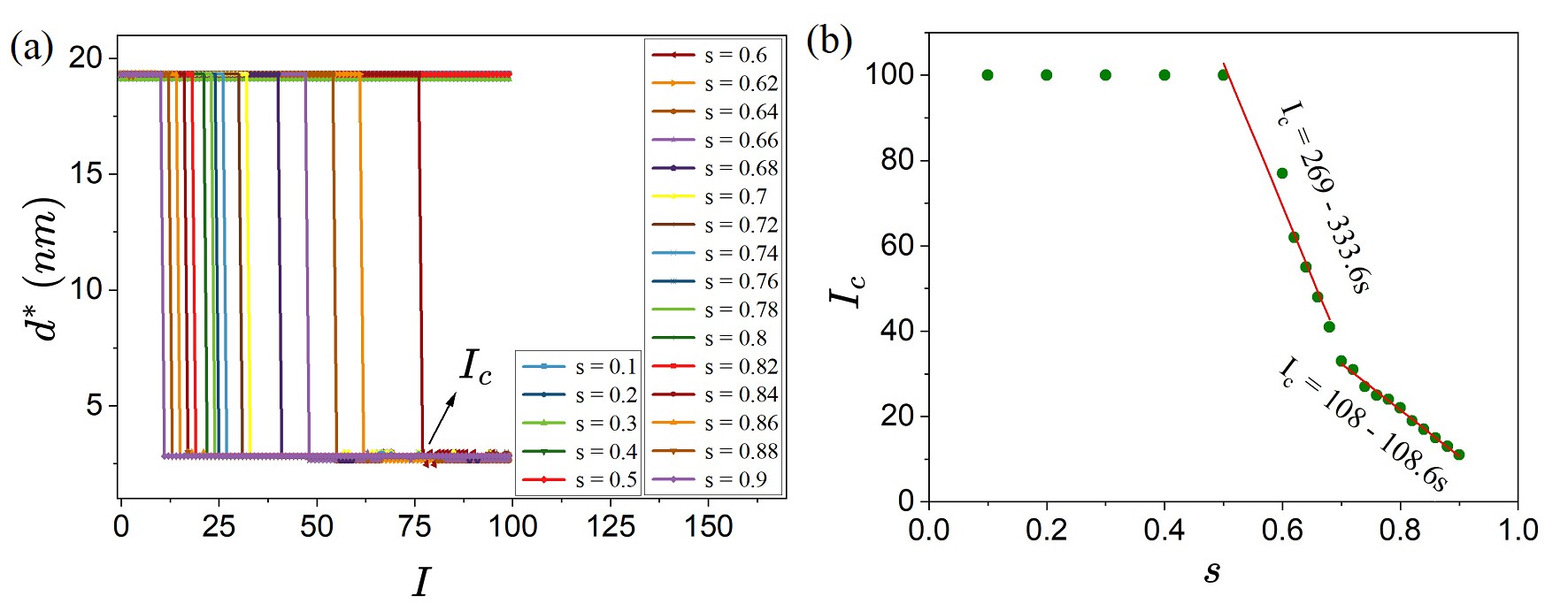} \ 
\caption{\footnotesize{ (a) Variation of the most frequently occurring particle size (represented by $d^*$) during the simulation with respect to $I$ for different diffusivity variables, indicating for small $s$ values, the system is dominated by particles of $\bar{d}\ =\ 18\ nm$. From $s\ =\ 0.6$, there exist a switch in the $d^*$ value with respect to different iterations. This implies that after a certain iteration (denoted as $I_c$) for a specific $s$, the maximum occurring $\bar{d}$ is approximately $3\ nm$. (b) The variation in iteration values at which the change in $d^*$ began (denoted as $I_c$) during the simulation for different diffusivity variables. The plot shows that the rate of change of $I_c$ with respect to $s$ is different for the two regions, indicated by different slope values for $s\ = 0.6$ to $0.68$ and $s \geq 0.7$ till $0.9$. For small $s$ values $I_c$ remains constant at 100. }}
\label{fig_5}
\end{figure*}

To gain more understanding of the behaviour observed in figure \ref{fig_2}(a), two parameters from the plot have been studied; (i) the maximum value of $\bar{b}_n$ for different $s$, and (ii) the corresponding value of iterations. Figure \ref{fig_2}(b) shows the maximum value of $\bar{b}_n$ indicated by $\bar{b}_n^*$ obtained for different diffusivity variables, and figure \ref{fig_2}(c) shows the corresponding value of iterations indicated by $I^*$. From figure \ref{fig_2}(b) and (c), it can be observed that for higher diffusivity variable, $\bar{b}_n^*$ has higher values which corresponds to the lesser iterations. This indicates that for higher values of $s$, i.e., beyond $s\ =\ 0.6$, more number of layers are breaking during the initiation of diffusion process itself. Whereas, for smaller $s$ values, even at 100th iteration, layers of the larger particles are not breaking, indicating that the strength of solvent is not enough to break the larger particles into smaller ones.

Figure \ref{fig_3}(a) represents the variation of average cumulative number of broken bonds ($\bar{b}_t$) with respect to iterations for different $s$. It can be seen from the plot that for small $s$, $\bar{b}_t$ does not exhibit much variation, even after a very high number of iterations. For intermediate $s$, $\bar{b}_t$ starts incrementing slowly with increasing number of iterations, experimentally indicating that with increase in reaction time bigger particles slowly start breaking into smaller particles. For very high values of $s$, $\bar{b}_t$ shows a higher initial value at the beginning of the simulation, which continues to increase further as $I$ increases. This indicates that, from the onset of diffusion, the solvent has sufficient strength to break the bonds of larger particles and form smaller ones. 

Figure \ref{fig_3}(b) further explains the results obtained for $\bar{b}_t$ vs $I$ by considering the $\bar{d}$ into account. $\bar{d}$ vs $\bar{b}_t$ shows an exponential decay for different iterations and for $s = 0.1$ to $0.9$, given by the equation,

\begin{equation} 
    \bar{d} = A e^{-m\bar{b}_t}
\end{equation}

where, $m$ represents the slope of the linear fit to the $ln(\bar{d})$ vs $\bar{b}_t$ plot, and $A$ represents the initial value of $\bar{d}$, corresponding to the intercept of the linear fit of  $ln(\bar{d})$ vs $\bar{b}_t$ plot. It is evident from the plot and the equation that $\bar{d}$ decreases with the increasing number of $\bar{b}_t$ and attains smaller values for higher $s$, which signifies that when the total number of bonds broken is increasing, the average size of the particles is reducing, consistent with the physical expectation. This behaviour has been examined for different values of $I$. It can be observed that this exponential behaviour is consistent for all the iterations (see inset). It should be noted that $I$ determines the extent to which the system evolves, as indicated by the elongation of the tail towards higher values of $\bar{b}_t$ when $I$ is increased. For increased $I$ values, higher $s$ results in more bonds being broken which consequently results in smaller $\bar{d}$.

\subsection{Fluctuation in the average size of nanoparticles and the attainment of saturation}

The size distributions of nanoparticles at different $s$ values, and different iterations have been shown in figure \ref{fig_4}. The figure represents size distribution for $s=0.1, 0.5$ and $0.9$, for three different iterations $10, \ 50$ and $100$. For $s=0.1$, during the entire diffusion process, the $\bar{d}$ remains around $20\ nm$. In contrast, for $s=0.9$, the size distribution indicates that the system contains a probability of smaller particles alongside larger ones, within very small values of $I$. As the number of iterations increases, the likelihood of smaller particles in the system also rises. This indicates that the solvent strength is strong enough to break more number of particles from the start of the diffusion process; in later iterations, the system has fewer particles of larger sizes than in earlier iterations. For intermediate values of $s$, the shift in the size distribution is observed after many iterations. Even at $I=100$ most of the particles are still around $20\ nm$ with only a few being smaller in average size. 
\begin{figure*}[ht]
\centering
\includegraphics[width=0.85\textwidth]{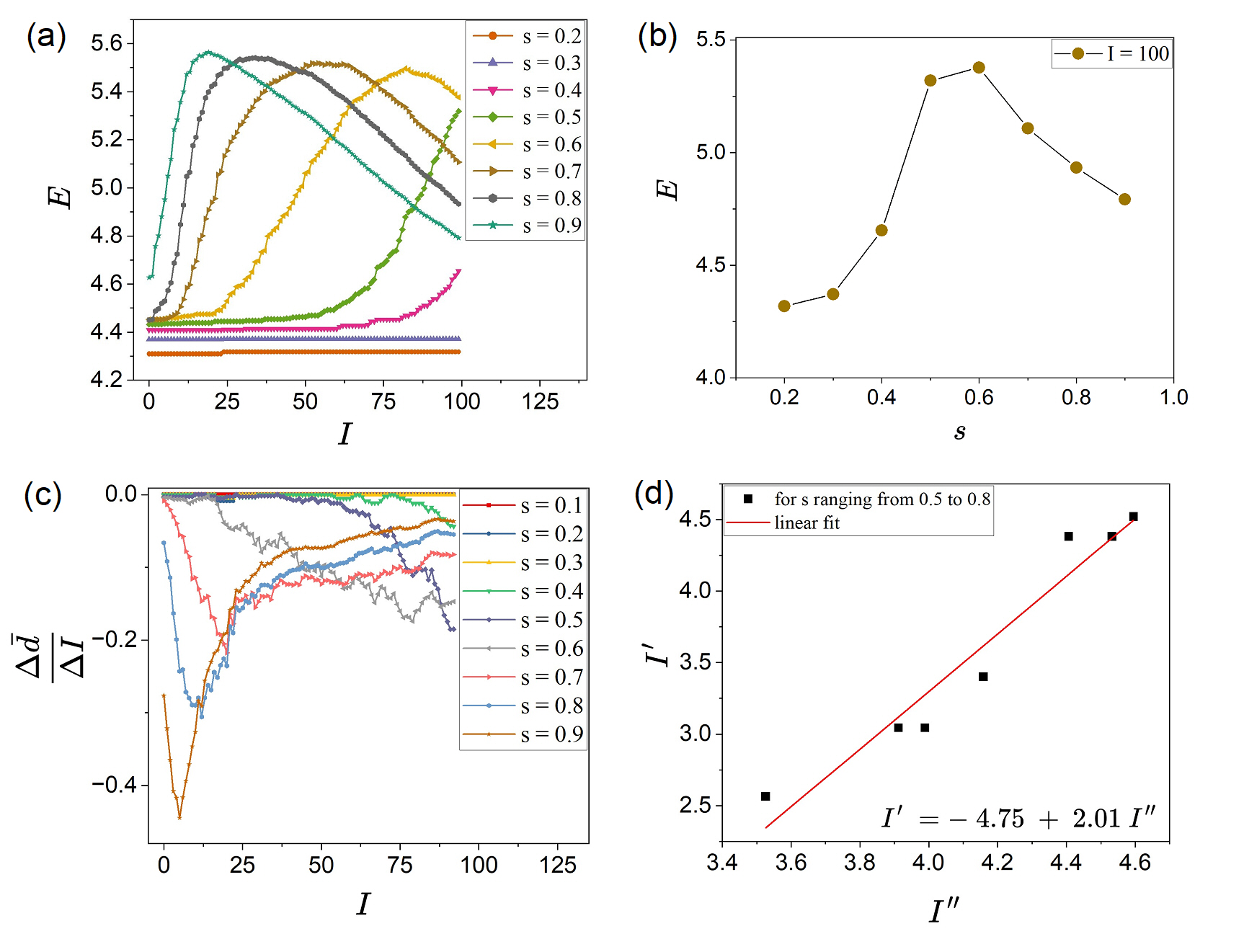} \\
\caption{\footnotesize{(a) Variation in entropy ($E$) with respect to iterations for different values of the diffusivity variable. The plot shows that $E$ remains unchanged for lower $s$ across all $I$ values. In contrast, intermediate $s$ values exhibit a gradual increment in $E$, and high $s$ values show a sharp increment up to a certain $I$ value. (b) Variation of entropy with respect to $s$ for $I\ =\ 100$. (c) The relative change of average particle size with iteration ($\frac{\Delta \bar{d}}{\Delta I}$) plotted against $I$ for all $s$ values, indicating that the $\frac{\Delta \bar{d}}{\Delta I}$ is more for higher $s$ values as compared to lower and intermediate $s$ values, which attains a minima at certain $I$ for different $s$ value, beyond which it starts reducing. (d) Correlation between the log of iteration corresponding to the maximum $E$ (denoted as I'') and log of iteration corresponding to the minimum of $\frac{\Delta \bar{d}}{\Delta I}$ (denoted as I') for $s$ ranging from 0.5 to 0.8.}}
\label{fig_6}
\end{figure*}

To understand the change in size distribution from higher to lower $\bar{d}$, a parameter called $d^*$ has been examined. This parameter indicates the most frequently occurring particle size in the system at each iteration throughout the simulation. $d^*$ has been studied for all $s$ values. Figure \ref{fig_5}(a) represents the variation of $d^*$ during the simulation, plotted against $I$ for different values of $s$. From the plot, it can be observed that the $d^*$ in the system is either around $3\ nm$ or $18\  nm$, with no other sizes occurring more frequently in between the range. This implies that during the evolution of system with iterations, even if there are particles with different $d$, the system will be dominated by particles with size $d^*$. The value of $d^*$ depends on the selection of $s$ and $I$. For example, for $s = 0.9$, during the first 40 iterations, the $d^*$ is $18\  nm$. After that, $d^*$ changes to $3\ nm$. 

The dependence of this switch in the $d^*$ on $I$ represented by $I_c$ for different $s$ values has been shown in figure \ref{fig_5}(b). The plot indicates that for $s=0.1$ to $0.5$, the switch in $d^*$ is not observed even at $I=100$. For $s$ values ranging from $0.6$ to $0.68$, the $I_c$ value sharply decreases linearly. This indicates that as $s$ increases, the number of iterations needed to observe the switching behavior decreases rapidly. In other words, for intermediate $s$, i.e., $0.6$ to $0.68$ the switch is being observed after a significant number of iterations as an effect of diffusion rate, resulting in higher value of the slope. The equation of line representing this behaviour is given as:
\begin{equation} 
    I_c = 269 - 333.6s 
\end{equation}

It should be noted that previously also (see figure \ref{fig_2}(c)) $I^*$ started showing reduction from $100$ only beyond $s\ =\ 0.6$. This indicates that the system begins transitioning from larger particles to smaller ones at a relatively less number of iterations as compared to $I\ =\ 100$, when $s$ exceeds 0.6. Additionally, the average number of new bonds broken i.e., $\bar{b}_n^*$ is also higher for values of $s$ exceeding 0.6 at the lower iteration values.
For $0.7$ to $0.9$, also the $I_c$ value reduces linearly with increase in s values, but with much smaller slope, with the equation given by
\begin{equation} 
    I_c = 108 - 108.6s 
\end{equation}

For higher $s$, i.e., $0.7$ to $0.9$ region the switch is observed within a very few iterations for consecutive $s$ values, resulting in smaller slope value. This occurs due to a very high diffusion rate, leading to a rapid shift in size distribution within just a few iterations after the diffusion process begins, as evident from figure \ref{fig_4}. From the distribution it can be visualized that for $s = 0.9$ and $I=10$, the probability  $P(\bar{d})$ is highest at $\bar{d} =\  18\  nm$. However, at $I\ =\ 50$, this maximum shifts to $\bar{d} =\ 3\ nm$. Also, from figure \ref{fig_5}(a), it can be understood that the $I$ value at which the switchover occurs for $d^*$ is different for different $s$ values.

To better quantify and understand the correlation between the size distribution plots for different $s$ at different $I$ values, Shannon entropy ($E$) of the system has been calculated using the formula \cite{ben17}:

\begin{equation} 
    E = -\sum_{\bar{d}=0}^{50} P(\bar{d}) \ln P(\bar{d}) 
\end{equation}

where, $P(\bar{d})$ is the probability of finding a particle of average size $\bar{d}$. Figure \ref{fig_6}(a) shows variation of entropy with respect to different iterations for different diffusivity variables. It can be observed that for small s, i.e. till $s\ =\ 0.3$, $E$ does not show any variation with $I$, indicating no change in the fluctuation of $\bar{d}$ for the particles in the system. For intermediate values of $s$ (specifically, $s = 0.4, 0.5$), a gradual increase in $E$ can be observed at the intermediate number of $I$. This indicates fluctuations in the average particle size within the system. It suggests that size non-uniformity is beginning to emerge due to the delamination of larger particles into smaller ones, which goes on increasing till $I=100$. For higher values of $s$, an abrupt increase in the system's entropy can be observed from the very initial points of $I$. This increase continues until it reaches a maximum at a specific value of $I$. Beyond this point, a decrease in $E$ can be noted. The initial increase in $E$ at very low $I$ values suggests that there is high fluctuation in the $\bar{d}$ values of the nanoparticles. This indicates that from the very beginning of the diffusion process, many larger particles have fragmented into smaller ones, resulting in the fluctuations of $E$ until a certain $I$ is reached. After this point, as fewer larger particles are available in the system, the particles exhibit reduced fluctuations in $\bar{d}$. This explains the onset of the reduction in $E$ with further increases in $I$. This reduction in $E$ with respect to $I$ can also be visualized from figure \ref{fig_4}, by comparing $I\ =\ 50$ and $100$ for $s\ =\ 0.9$.

Figure \ref{fig_6}(b) represents the variation of $E$ with $s$ for 100th iteration. It can be observed that entropy shows non-monotonic behaviour with respect to $s$ for the 100th iteration. $E$ is increasing for smaller $s$ values, which attains a maximum for intermediate $s$, and then reduces for higher $s$. This plot again indicates how the systems are evolving depending on $I$. For smaller $s$, less fluctuation at maximum $I$ considered indicates that the system has not evolved much from the initial state. Slightly increased fluctuation is observed for $s\ =\ 0.4$, indicating the onset of non-uniformity in the $\bar{d}$ of the nanoparticles. On contrary the intermediate $s\ =\ 0.5,\ 0.6$ values show maximum entropy as compared to other $s$ values, which indicates increased non-uniformity during the evolution of systems with respect to $I$. Entropy of the system starts reducing beyond $s\ =\ 0.6$. This reduction for higher $s$ values indicates that, with further increase in $s$ value the systems are attaining more uniformity in terms of $\bar{d}$. The recurrence of $s\ =\ 0.6$ is noteworthy, as previously mentioned for figures \ref{fig_2}(c) and \ref{fig_5}(b). This value indicates the optimal $s$ beyond which the system shows relatively abrupt transitions in particle size reduction compared to smaller and intermediate $s$ values. This behavior is evident from the observations made regarding the iterations of $\bar{b}_n^*$, $d^*$, as well as in terms of $E$ where more size uniformity for the smaller nanoparticles is observed. The three regions in the figure \ref{fig_6}(b) indicates different stages of evolution, which depends significantly on number of iterations; (i) For small $s$ values, since the solvent does not have enough strength to delaminate the layers of nanoparticles, no fluctuation has been observed even at 100th iteration. (ii) For intermediate $s$, the higher $E$ value suggests the system is evolving, and is at the stage where particles of different sizes are present in the system. (iii) For higher $s$, since the diffusion rate is very high, the system has evolved quickly and has particles of much smaller sizes in the system as evident from figure \ref{fig_4}.

Figure \ref{fig_6}(c) shows the relative change of average particle size with iteration represented as $\frac{\Delta \bar{d}}{\Delta I}$, which is plotted against $I$ for different $s$ values. Higher the difference between the $\bar{d}$ values for consecutive iterations, the more negative rate of change will be with respect to $I$. As shown in figure \ref{fig_1}, for small values of $s$, $\bar{d}$ exhibits no variation throughout the diffusion process; consequently, Figure \ref{fig_6}(c) shows no deviation from zero. In the case of higher $s$ values, for small values of $I$, $\frac{\Delta \bar{d}}{\Delta I}$ starts showing decrement and attains a minimum value. Beyond this minimum value it starts to show increment again and gradually attains saturation with high $I$ values. In this case the initial decrement represents the abruptness in reduction of $\bar{d}$ of the particles until the minimum value is attained. Beyond the minimum value, $\frac{\Delta \bar{d}}{\Delta I}$ starts showing increment towards zero since the non-uniformity in the $\bar{d}$ starts to reduce after a few iterations, slowly taking the system towards saturation. For intermediate values of $s$, the reduction towards the minimum value appears to be delayed. This delay occurs because the decrease in the non-uniformity of $\bar{d}$ for these intermediate $s$ values is only achieved after many more iterations, specifically beyond $I\ =\ 100$. This indicates that for intermediate $s$ values, a minimum may not have been reached by the time $I\ =\ 100$, as discussed for figure \ref{fig_6}(b).

From figure \ref{fig_6}(a) and (c), the system can be observed to move towards saturation in terms of uniformity in average size of the particles, after reaching a specific $I$ value for different $s$. To further understand this behaviour, the log of iterations corresponding to the maximum value obtained in \ref{fig_6}(a) denoted as $I''$ has been plotted along with the iteration corresponding to the minimum obtained in \ref{fig_6}(c) represented by $I'$, for the $s$ ranging from 0.5 to 0.8. For $s<0.5$ the system does not show much variation in $\bar{d}$, even for the maximum iterations considered during the simulation. Conversely, for $s=0.9$ , the system experiences an abrupt variation in $\bar{d}$ at the start of the diffusion process, complicating the analysis of particle size evolution in the system. By considering these two cases as extremes, range of this analysis has been restricted to $s\ =\ 0.5$ to $0.8$. Figure \ref{fig_6}(d) allows us to understand the existence or non-existence of the correlation in between the two parameters. From figure \ref{fig_6}(d), it is evident that these two parameters show power law behaviour with an exponent $2.01$, and the linear fit indicates that there exist a correlation in between the iterations corresponding to the maximum value of entropy and iterations corresponding to the minimum value of $\frac{\Delta \bar{d}}{\Delta I}$ in the range 0.5 to 0.8. This correlation signifies that there exists an iteration value which corresponds to take the system towards more uniformity in terms of $\bar{d}$ of the nanoparticles. Beyond these specific iteration values for different $s$, the system slowly starts moving towards attainment of saturation in terms of uniformity in $\bar{d}$.

\section{Conclusion}

This study presents the numerical simulation of solvent diffusion in transition metal dichalcogenides based-nanomaterials in solvothermal reactions, leading to the exfoliation of layers and hence the reduction in the average size of the nanoparticles. By solving modified Fick’s law of diffusion and utilizing the dynamic bond percolation model, the evolution of a system of nanoparticles has been studied. The diffusion process has been realized by varying the  parameters such as diffusivity variable and the number of iterations taken during the simulation, along with some disordeness in the material property. Here, diffusivity is the parameter which depends on the selection of solvent during the solvothermal reaction or it will also depend upon the reaction parameters for a particular solvent. The number of iteration in the present study represents the parameter similar to reaction time considered during the reaction. 

The results indicate that diffusivity variable and the number of iterations considered during the diffusion process play very important role in determining the average size of the particles in the system. For a range of diffusivity variables considered in this study, the solvents with lower values of diffusivity variable do not show any variation in the average size of the nanoparticles until the maximum number of iteration is reached. On the contrary, the solvents with higher diffusivity variable changes the average size of the particles within few number of iterations. This gives information about the importance of strength of the solvent taken for the delamination of larger nanoparticles into smaller ones in the solvothermal reaction. This is evidenced by the elaborative study of the avalanche statistics as well as fluctuation observed in the average size through the entropy calculations. There is still future scope to understand and incorporate surface energy into the numerical simulations, as it is a crucial parameter for nano-regime systems. This can further improve our understanding of the growth dynamics in solvothermal-related systems.

\section{Ackowledgement}

A special thank goes to Prof. Souri Banerjee and Dr. Subhadeep Roy for their guidance throughout the work. I also want to thank Shyamapada Pal, Anjali Vajigi and Viswakannan R.K. for useful discussions.




\begin{thebibliography}{99}

\bibitem{manzeli17} Manzeli, Sajedeh, Dmitry Ovchinnikov, Diego Pasquier, Oleg V. Yazyev, and Andras Kis. "2D transition metal dichalcogenides." Nature Reviews Materials 2, no. 8 (2017): 17033.
\bibitem{kumar25} Kumar, Raj, Chandrani Sarkar, Naveen Bunekar, Yogendra Kumar Mishra, and Ajeet Kaushik. "State-of-the-art transition metal dichalcogenides: Synthesis, functionalization, and biomedical applications." Materials Today (2025).
\bibitem{Cao18} Cao, Xuanyu, Caiping Ding, Cuiling Zhang, Wei Gu, Yinghan Yan, Xinhao Shi, and Yuezhong Xian. "Transition metal dichalcogenide quantum dots: synthesis, photoluminescence and biological applications." Journal of materials chemistry B 6, no. 48 (2018): 8011-8036.
\bibitem{uddin25} Uddin, Mohsin, Sumaiya Sumaiya, Syed Muhammad Osama, and Syed Hassan Abbas. "Emerging trends in 2D materials: Beyond graphene for next-generation applications." Mechanics Exploration and Material Innovation 2, no. 1 (2025): 38-48.
\bibitem{emil06} Roduner, Emil. "Size matters: why nanomaterials are different." Chemical society reviews 35, no. 7 (2006): 583-592.
\bibitem{mansur10} Mansur, Herman Sander. "Quantum dots and nanocomposites." Wiley Interdisciplinary Reviews: Nanomedicine and Nanobiotechnology 2, no. 2 (2010): 113-129.
\bibitem{pratap25} Pratap, Surya. "A review on development in synthesis process, characteristics, and biological applications of transition metal dichalcogenides quantum dots." Tissue and Cell (2025): 103156.
\bibitem{haz22} Abdelsalam, Hazem, and Qin Fang Zhang. "Properties and applications of quantum dots derived from two-dimensional materials." Advances in Physics: X 7, no. 1 (2022): 2048966.
\bibitem{sahu26} Sahu, Geetika, Alle Pawan Kumar Reddy, R. K. Viswakannan, Aritra Chatterjee, Subhadeep Roy, and Souri Banerjee. "Experimental and Numerical Optimization of Mechanical Properties of PVA Nanocomposites With Molybdenum Disulfide Quantum Dots." Polymers for Advanced Technologies 37, no. 1 (2026): e70494.
\bibitem{jana09} Drbohlavova, Jana, Vojtech Adam, Rene Kizek, and Jaromir Hubalek. "Quantum dots—characterization, preparation and usage in biological systems." International journal of molecular sciences 10, no. 2 (2009): 656-673.
\bibitem{Zh15} Zhang, Jianbing, Ryan W. Crisp, Jianbo Gao, Daniel M. Kroupa, Matthew C. Beard, and Joseph M. Luther. "Synthetic conditions for high-accuracy size control of PbS quantum dots." The journal of physical chemistry letters 6, no. 10 (2015): 1830-1833.
\bibitem{mehrer2007} Mehrer, Helmut. "Dependence of diffusion on temperature and pressure." Diffusion in solids: Fundamentals, methods, materials, diffusion-controlled processes (2007): 127-149.
\bibitem{sahu24} Sahu, Geetika, Chanchal Chakraborty, Subhadeep Roy, and Souri Banerjee. "Dependence of reaction time in hydrothermal synthesis of MoS2 quantum dots: An investigation using optical tools and fractal analysis." Journal of Crystal Growth 627 (2024): 127487.
\bibitem{sahu25} Sahu, Geetika, Chanchal Chakraborty, Subhadeep Roy, and Souri Banerjee. "Non-monotonic growth of MoS2 quantum dots observed through fractal analysis: role of precursor concentration on the overall growth dynamics." Physica Scripta 100, no. 5 (2025): 055931.
\bibitem{becker11} Becker, John, Krishna Reddy Raghupathi, Jordan St. Pierre, Dan Zhao, and Ranjit T. Koodali. "Tuning of the crystallite and particle sizes of ZnO nanocrystalline materials in solvothermal synthesis and their photocatalytic activity for dye degradation." The Journal of Physical Chemistry C 115, no. 28 (2011): 13844-13850.
\bibitem{rama18} Ramakrishnan, Venkatraman Madurai, Muthukumarasamy Natarajan, Agilan Santhanam, Vijayshankar Asokan, and Dhayalan Velauthapillai. "Size controlled synthesis of TiO2 nanoparticles by modified solvothermal method towards effective photo catalytic and photovoltaic applications." Materials Research Bulletin 97 (2018): 351-360.
\bibitem{dem08} Demazeau, Gérard. "Solvothermal reactions: an original route for the synthesis of novel materials." Journal of Materials Science 43, no. 7 (2008): 2104-2114.
\bibitem{kov21} Kovács, Zoltán, Csanád Molnár, Urška Lavrenčič Štangar, Vasile-Mircea Cristea, Zsolt Pap, Klara Hernadi, and Lucian Baia. "Optimization method of the solvothermal parameters using Box–Behnken experimental Design—The case study of ZnO structural and catalytic tailoring." Nanomaterials 11, no. 5 (2021): 1334.
\bibitem{harris86} Harris, Caroline S., A. Nitzan, Mark A. Ratner, and D. F. Shriver. "Particle motion through a dynamically disordered medium: The effects of bond correlation and application to polymer solid electrolytes." Solid State Ionics 18 (1986): 151-155.
\bibitem{ali22} Ali, Luqman, Fazle Subhan, Muhammad Ayaz, Syed Shams ul Hassan, Clare Chisu Byeon, Jong Su Kim, and Simona Bungau. "Exfoliation of MoS2 quantum dots: recent progress and challenges." Nanomaterials 12, no. 19 (2022): 3465.
\bibitem{wang22} Wang, Qi. "Mechanism investigation of synthesizing carbon quantum dots from coal by numerical simulation." PhD diss., 2022.
\bibitem{druger83} Druger, Stephen D., Abraham Nitzan, and Mark A. Ratner. "Dynamic bond percolation theory: A microscopic model for diffusion in dynamically disordered systems. I. Definition and one‐dimensional case." The Journal of chemical physics 79, no. 6 (1983): 3133-3142.
\bibitem{samy21} Samy, Omnia, Shuwen Zeng, Muhammad Danang Birowosuto, and Amine El Moutaouakil. "A review on MoS2 properties, synthesis, sensing applications and challenges." Crystals 11, no. 4 (2021): 355.
\bibitem{guo20} Guo, Yongming, and Jianwei Li. "MoS2 quantum dots: synthesis, properties and biological applications." Materials Science and Engineering: C 109 (2020): 110511.
\bibitem{wang22} Wang, Qi. "Mechanism investigation of synthesizing carbon quantum dots from coal by numerical simulation." PhD diss., 2022.
\bibitem{xu92} Xu, Gu. "The effects of chain segment motion on ionic diffusion in solid polymer electrolytes." Solid state ionics 50, no. 3-4 (1992): 345-347.
\bibitem{Lee05} Lee, Jin-Seok, and Sung-Churl Choi. "Solvent effect on synthesis of indium tin oxide nano-powders by a solvothermal process." Journal of the European Ceramic Society 25, no. 14 (2005): 3307-3314.
\bibitem{chan14} Chan, T. C., Irene Lee, and K. S. Chan. "Effect of solvent on diffusion: probing with nonpolar solutes." The Journal of Physical Chemistry B 118, no. 37 (2014): 10945-10955.
\bibitem{rahaman15} Rahaman, M. M., M. Sikdar, M. B. Hossain, M. Rahaman, and M. J. Hossain. "Numerical Solution of Diffusion Equation by Finite Difference Method." IOSR J. Math. 11 (2015): 2278-5728.
\bibitem{john18} John, Bincy, G. Genifer Silvena, and A. Leo Rajesh. "Influence of reaction time on the structural, optical and electrical performance of copper antimony sulfide nanoparticles using solvothermal method." Physica B: Condensed Matter 537 (2018): 243-250.
\bibitem{wu25} Wu, Shunnian, WP Cathie Lee, Hashan N. Thenuwara, and Ping Wu. "Quantitative criteria for solvent selection in Liquid-Phase exfoliation: balancing exfoliation and stabilization efficiency." Nanomaterials 15, no. 5 (2025): 370.
\bibitem{coleman13} Coleman, Jonathan N. "Liquid exfoliation of defect-free graphene." Accounts of chemical research 46, no. 1 (2013): 14-22.
\bibitem{loh15} Loh, G. C., Ravindra Pandey, Yoke Khin Yap, and Shashi P. Karna. "MoS2 quantum dot: Effects of passivation, additional layer, and h-BN substrate on its stability and electronic properties." The Journal of Physical Chemistry C 119, no. 3 (2015): 1565-1574.
\bibitem{abr22}Abreu, Barbara, Bernardo Almeida, Pedro Ferreira, Ricardo MF Fernandes, Diana M. Fernandes, and Eduardo F. Marques. "A critical assessment of the role of ionic surfactants in the exfoliation and stabilization of 2D nanosheets: The case of the transition metal dichalcogenides MoS2, WS2 and MoSe2." Journal of colloid and interface science 626 (2022): 167-177.
\bibitem{ben17} Ben-Naim, Arieh. "Entropy, Shannon’s measure of information and Boltzmann’s H-theorem." Entropy 19, no. 2 (2017): 48.

\end{thebibliography}
\end{document}